\documentclass{article}

\usepackage{arxiv}

\usepackage[utf8]{inputenc}
\usepackage[T1]{fontenc}
\usepackage{hyperref}
\usepackage{url}
\usepackage{booktabs}
\usepackage{amsmath, amssymb, amsfonts}
\usepackage{nicefrac}
\usepackage{microtype}
\usepackage{graphicx}
\usepackage{natbib}
\usepackage{doi}

\usepackage{float}
\usepackage{multirow}   

\usepackage{subcaption} 

\title{Assignment–Routing Optimization with Cutting-Plane Subtour Elimination: Solver and Benchmark Dataset}

\author{
  {\includegraphics[scale=0.06]{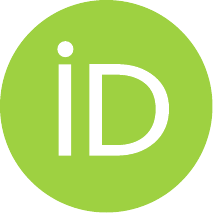}\hspace{1mm} Yuan Qilong } \\
  Singapore Institute of Technology \\  \texttt{qilong.yuan@singaporetech.edu.sg} \\
}

\hypersetup{
  pdftitle={Assignment–Routing Optimization with Cutting-Plane Subtour Elimination: Solver and Benchmark Dataset},
  pdfsubject={Optimization, Combinatorics, Gurobi},
  pdfauthor={Yuan Qilong},
  pdfkeywords={Joint Routing–Assignment Optimization, Mixed Integer Programming , Cutting-Plane Method, Datasets},
}

\begin{document}
\maketitle

\begin{abstract}
We study a joint routing–assignment optimization problem in which a set of items must be paired one-to-one with a set of placeholders while simultaneously determining a Hamiltonian cycle that visits every node exactly once. Both the assignment and routing decisions are optimized jointly to minimize the total travel cost. In this work, we propose a method to solve this problem using an exact MIP formulation with Gurobi, including cutting-plane subtour elimination. With analysis of the computational complexity and through extensive experiments, we analyze the computational limitations of this approach as the problem size grows and reveal the challenges associated with the need for more efficient algorithms for larger instances. The dataset, formulations, and experimental results provided here can serve as benchmarks for future studies in this research area. Many pickup-placing with routing problems can be solved with such method with addition assignment constraints. GitHub repository : \url{https://github.com/QL-YUAN/Joint-Assignment-Routing-Optimization}
\end{abstract}

\keywords{Joint Routing–Assignment Optimization \and Mixed Integer Programming \and Cutting-Plane Method \and Datasets}

\section{Introduction}
\label{sec:intro}

Many practical tasks in robotics, logistics, and manufacturing require transporting a collection of discrete objects from their current locations to designated target places. 
Such problems naturally couple three interdependent decision layers: 
(i) \emph{assignment} — determining which object should go to which labeled placeholder; 
(ii) \emph{sequencing} — deciding in what order the agent should pick up and deliver the objects; and 
(iii) \emph{routing or motion planning} — finding a feasible path that minimizes travel cost while satisfying geometric and kinematic. This paper focus on first two concerns.

A range of realistic scenarios are directly related:

\begin{itemize}
  \item \textbf{Vehicle-based pickup and drop-off.} An autonomous vehicle or courier robot must pick up multiple passengers or packages (one at a time) and deliver each to a unique destination. Determining the optimal pickup–drop-off pairings and the overall route forms a cyclic tour that minimizes travel cost and maximizes efficiency.

  \item \textbf{Tabletop rearrangement with robot arms.} A robotic manipulator moves scattered items (e.g., puzzle pieces or tools) into specific target positions on a workspace. Each item must be assigned to a placeholder, and the robot must plan sequence that visits all pick-and-place locations efficiently.

  \item \textbf{Automated packaging and assembly.} A robotic or human operator sequentially transfers items (e.g., pastries, gadgets, or components) from a source area into predefined slots in packaging trays or boxes. The system must determine both which item goes where and in what order to perform the transfers to reduce time or effort.

  \item \textbf{Service robotics and delivery logistics.} A service robot performs repeated pick-and-place tasks, such as collecting food orders or mail items and delivering each to a designated locker or customer. Optimizing the task sequence and assignments jointly improves response time and route efficiency.
\end{itemize}

Although compact in statement, this problem integrates the combinatorial structure of assignment and routing with precedence and connectivity constraints. 
The decision variables include a bijective mapping and a tour permutation, while the agent must obey unit-capacity and pickup-before-delivery constraints. 
The objective is typically to minimize total travel distance in this work, but extensions may incorporate other costs such manipulation costs, stopping overheads etc. which could be considered in futureworks.

\subsection{Connections to existing problem classes}
This problem lies at the intersection of several established families in operations research and robotics.

\paragraph{Pickup-and-delivery and the stacker-crane problem.}
Pickup-and-delivery variants of the traveling salesman and vehicle routing problems (PDTSP, VRP-PD) enforce precedence constraints between pickup and delivery pairs under vehicle capacity limits. 
The stacker-crane abstraction \cite{hernandez2004,hernandez2007,treleaven2012} is closely related, representing a single-vehicle system that must transport items between paired pickup and drop-off locations. 
These problems are NP-hard in general and have motivated a rich literature on branch-and-cut formulations, exact algorithms for small-scale instances, and heuristic or metaheuristic methods for larger ones. 

\paragraph{Assignment–routing integration.}
Classical formulations of the assignment problem \cite{kuhn1955hungarian} and the traveling salesman problem \cite{dantzig1954solution, lawler1985traveling, applegate2006traveling} address individual subcomponents of our task. 
However, when assignment and routing are coupled the search space expands combinatorially, requiring integrated formulations or hierarchical decomposition. 
Recent robotics literature has explored mixed-integer and sampling-based approaches for such coupled problems, yet scalability and global optimality remain open challenges. 
Modern solvers such as Gurobi \cite{gurobi2024manual} allow constraint families to be activated dynamically through cutting-plane method based callbacks, a strategy that has been successfully employed in subtour elimination for TSPs and connectivity-enforced routing \cite{applegate2006traveling}. 
This mechanism reduces model size and computational overhead by adding connectivity constraints only when violated, which we exploit in our implementation. 

\subsection{Purposes of the Article}
We present a formulation that explicitly combines assignment and routing with precedence and connectivity constraints, forming a single Hamiltonian cycle over all pickup and delivery nodes. 

The main purposes of this paper are:
\begin{enumerate}
  \item A compact mixed-integer programming (MIP) model that jointly incorporates assignment, capacity, and cycle constraints, with an efficient Gurobi implementation using cutting-plane constraints to dynamically eliminate subtours.
  \item Publicly available benchmark datasets, including input data and computed optimal solutions, to facilitate reproducibility and serve as a reference for future research.
  \item Analysis of the computational complexity of this MIP formulation, highlighting the limitations of exact solution methods and motivating the development of more efficient heuristic algorithms.
\end{enumerate}

\section{Mathematical Formulation}

Let $I = \{0, \dots, n-1\}$ be the set of items, $P = \{n, \dots, 2n-1\}$ the set of placeholders, and $c_{ij}$ the cost of traversing edge $(i,j)$. Denote $V = I \cup P$ and $E = \{(i,j) \in V \times V : i \neq j\}$. 

We define binary decision variables:
\[
x_{ij} =
\begin{cases}
1, & \text{if edge } (i,j) \text{ is included in the cycle}, \\[4pt]
0, & \text{otherwise.}
\end{cases}
\]

The objective is to minimize total traversal cost:
\[
\min \sum_{(i,j) \in E} c_{ij} x_{ij}.
\]
\subsection{Simplified Model Constraints}

In order to solve the problem efficiently, in this problem modeling, only edges from item to placeholder is allowed.  Constraints are as follows:
\begin{enumerate}
    \item \textbf{Valid edges:} In order to make the model as simple as possible for high high computational efficiency , here only item-to-placeholder edges are allowed. Since straight forward post processing of edges can result in complete cycle with directions, such definition is fine for standard assignment and routing problem. The valid edges constraints are as follows:
    \[
    x_{ij} = 0 \text{ if both } i,j \in I \text{ or } i,j \in P \text{ or } ( i \in P  \text{ and } j \in I)
    \]
    \item \textbf{Assignment:} Each item and placeholder is assigned twice:
    \[
    \sum_{p \in P} x_{ip} = 2, \quad \forall i \in I, 
    \quad
    \sum_{i \in I} x_{ip} = 2, \quad \forall p \in P
    \]
    \item \textbf{Fixed pair:} In case of having starting and goal points, we specify starting position to be the last placeholder, goal point to be the last item, and enforce the last item to be assigned to the last placeholder as following constraints. 

\[
x_{n-1, 2n-1} = 1,x_{2n-1, n-1} = 0
\]
    \item \textbf{Subtour elimination:} Disconnected cycles are prohibited by a cutting-plane method \cite{kelley1960cutting,westerlund1995extended}:
    \[
    \sum_{i,j \in S, i \neq j} x_{ij} \le |S|-1, \quad \forall \text{ subtours } S \subset V
    \]
    This is enforced dynamically using subtour elimination callback function in Gurobi.
\end{enumerate}

\subsection{Generalization of Modeling for Problem with More Constraints}

The formulation introduced above is suitable for the case of a single type of items and placeholders, where each item can be assigned to any placeholder. However, many practical applications involve multiple types or categories of items, each associated with a distinct subset of compatible placeholders. Examples include multi-type packing tasks, board or game piece arrangements, and robotic placement operations where items of the same type are interchangeable but must be placed in designated regions.

We will have follow-up articals to discuss on methods to solve such problems. However, first of all, in such scenarios, the simplified model as mentioned above will have difficulty defining the constraints. In such cases we would like the edges to be directional instead of only connecting from item to placeholder. At the same time, we need to introduce additional binary decision variables:
\[
a_{ip} = \begin{cases} 1 & \text{if item } i \text{ is assigned to placeholder } p \\ 0 & \text{otherwise} \end{cases}.
\]
Then, the constraints of the joint assignment-routing problem becomes:
\begin{enumerate}
    \item \textbf{Degree:} Each node has exactly two incident edges:
    \[
    \sum_{j \in V, j \neq i} x_{ij} + \sum_{j \in V, j \neq i} x_{ji} = 2, \quad \forall i \in V
    \]
    \item \textbf{Assignment:} Each item and placeholder is assigned twice:
    \[
    \sum_{p \in P} a_{ip} = 2, \quad \forall i \in I, 
    \quad
    \sum_{i \in I} a_{ip} = 2, \quad \forall p \in P
    \]
    \item \textbf{Valid edges:} Only item-to-placeholder edges are allowed, and edges can only exist if the assignment exists (here placeholder to item edges is allowed in this modeling):
    \[
    x_{ij} = 0 \text{ if both } i,j \in I \text{ or } i,j \in P, 
    \quad
    x_{ip} \le a_{ip}, \quad x_{pi} \le a_{ip}, \forall i \in I, p \in P
    \]
    \item \textbf{Fixed pair:} In case of having starting and goal points, we specify starting position to be the last placeholder, goal point to be the last item, and enforce the last item to be assigned to the last placeholder as following constraints. 
\[
a_{n-1, 2n-1} = 1
\]

\[
x_{n-1, 2n-1} = 1,x_{2n-1, n-1} = 0
\]
    \item \textbf{Subtour elimination:} Disconnected cycles are prohibited by a cutting-plane method:
    \[
    \sum_{i,j \in S, i \neq j} x_{ij} \le |S|-1, \quad \forall \text{ subtours } S \subset V
    \]
    \item Following constraints need to be added to speed up the solver. 
    \[
    x_{ij} + x_{ji}<=1,  \forall i, \forall j,  i \neq j
    \]
\end{enumerate}

\section{Computational Complexity}

The pure assignment problem can be solved in polynomial time 
using the Hungarian algorithm \cite{kuhn1955hungarian} in $O(n^3)$ complexity. 
However, the present formulation couples the assignment decision 
with routing optimization, the \emph{Traveling Salesman Problem} (TSP) with fixed pick-place tasks constraints with complexity of $O(n!)$, making the problem significantly harder in $O(n!*n^3)$ complexity.

\subsection{Computational Complexity of The Proposed MIP Solver}

The mixed-integer programming (MIP) solver in Gurobi employ a branch-and-bound framework augmented with cutting planes, preprocessing, heuristics and node presolve (commonly referred to as branch-and-cut).
Specifically, the model requires finding both:
(i) a bijective assignment between items and placeholders, and 
(ii) a single traversal order (tour) that visits all nodes exactly once 
while satisfying assignment consistency and pickup–delivery precedence constraints.

The MIP model contains the following variables and constraints:
\begin{itemize}
    \item $O(n^2)$ binary routing variables $x_{ij}$, and $O(n^2)$ linear constraints.
    \item An exponential number of subtour elimination constraints introduced dynamically by the cutting-plane procedure \cite{basu2020complexitybranchandboundcuttingplanes}.
\end{itemize}

Let \(m\) denote the number of binary decision variables in the MIP. For our specific formulation, the dominant binary variables are the routing and assignment binaries: $m = \Theta(n^2),$

At each search instance the solver typically (i) solves a linear programming (LP) relaxation and (ii) runs presolve / separation routines. Let \(T_{\text{LP}}(s)\) denote the time to solve an LP of size \(s\) (which is $O(2^m)$) and \(T_{\text{sep}}(s)\) denotes the time spent on all separation/cut generation at the instance. LP algorithms are proved to be polynomial in \(s\) \cite{karmarkar1984new}, so we can write \(T_{\text{LP}}(s)=\operatorname{poly}(s)\). However $T_{\text{sep}(s)}$ is exponential function of $s$ in worse-case, according to study in \cite{basu2020complexitybranchandboundcuttingplanes}. In other words, in worse-case, the MIP algorithm has \textbf{exponential worst-case time complexity}. Thus, the worst-case computational effort grows exponentially with problem size ($n$), 
and only small to medium instances can be solved to proven optimality 
using current MIP solvers \cite{basu2020complexitybranchandboundcuttingplanes,gurobi2025mipprimer}.
\section{Datasets}
All the datasets files can be downloaded from project GitHub repository. The experiments presented here applied the simplified model as introduced in Section 2.

\noindent The following Table  \ref{tab:datasets} shows a sample of data files in the datasets used for experiments. Here, \textbf{Experiment} is the experiment ID (starting from 1000), \textbf{ID} is the item/ placeholder index, \textbf{pX}, \textbf{pY} are the coordinates of the item, and \textbf{tX}, \textbf{tY} are the coordinates of the placeholders.

\begin{table}[h!]
\footnotesize
\centering
\caption{Item and placeholder coordinates for Experiment 1000}
\label{tab:datasets}
\begin{tabular}{cccccc}
\toprule
\textbf{Experiment} & \textbf{ID} & \textbf{pX} & \textbf{pY} & \textbf{tX} & \textbf{tY} \\
\midrule
1000 & 0 & 0.521386 & 0.603842 & 0.974764 & 0.153932 \\
1000 & 1 & 0.470942 & 0.203248 & 0.699089 & 0.447241 \\
1000 & 2 & 0.528759 & 0.191036 & 0.017513 & 0.291025 \\
\bottomrule
\end{tabular}
\end{table}

All experiments were conducted on a workstation equipped with an Intel(R) Core(TM) Ultra 7 155H processor, supporting the SSE2, AVX, and AVX2 instruction sets. The CPU has 22 physical cores and 22 logical processors, and all experiments were configured to utilize up to 22 threads for parallel computations.
\noindent The following table lists the number of node pairs and the corresponding data files used for experiments. The node coordinates for each experiment are stored in CSV files named as \textbf{experimental\_n\_\{$n$\}\_data.csv}, as shown in Table \ref{tab:files}.

\begin{table}[h!]
\footnotesize
\centering
\caption{Number of node pairs and corresponding data files}
\label{tab:files}
\begin{tabular}{cc}
\toprule
Number of Node Pairs ($n$) & File Name \\
\midrule
10   & experimental\_n\_10\_data.csv \\
20   & experimental\_n\_20\_data.csv \\
32*   & experimental\_n\_32*\_data.csv \\
100  & experimental\_n\_100\_data.csv \\
200  & experimental\_n\_200\_data.csv \\
300  & experimental\_n\_300\_data.csv \\
400  & experimental\_n\_400\_data.csv \\
500  & experimental\_n\_500\_data.csv \\
1000 & experimental\_n\_1000\_data.csv \\
\bottomrule
\end{tabular}
\end{table}

\subsection*{Experiments and Results Files}
All experiments were performed using the \textbf{Gurobi Optimizer} with a maximum runtime (timeout) set to sufficient time for completing each instance (When n=100 or smaller, 5mins is more than enough. However, when n=300, instances can easily take more than 10mins). \textit{Source code to run the experiment can be found in GitHub following the readme instruction}.  The solver was configured to utilize all available threads on the workstation, enabling parallel execution of the algorithm.

\paragraph{Tested dataset samples:}  
The following dataset samples  were evaluated with Gurobi. The corresponding results are stored in CSV files named as \textbf{updated\_experiment\_n\_\{$n$\}\_results.csv}, and the related images are stored in folders named \textbf{img\_update\{$n$\}}, as shown in Table \ref{tab:result_files}.

Table \ref{tab:results} provides a sample of the experimental results with each row being the result of one single experiment:

\begin{itemize}
    \item \textbf{experiment\_id}: Unique identifier of the experiment.
    \item \textbf{best\_cost}: Best objective value found by Gurobi, which is the optimal solution cost.
    \item \textbf{dt}: Total computation time in seconds. This should be within timeout settings. 
    \item \textbf{edges}: List of edges used in the solution (e.g., for routing problems).
\end{itemize}

\begin{table}[h!]
\footnotesize
\centering
\caption{Tested problem instances, result files, and corresponding image folders}
\label{tab:result_files}
\begin{tabular}{ccc}
\toprule
Instance & Result File & Image Folder \\
\midrule
$n=32\_ic$ & updated\_experiment\_n\_32\_ic\_results.csv & img\_update\_32\_ic \\
$n=32\_xq$ & updated\_experiment\_n\_32\_xq\_results.csv & img\_update\_32\_xq \\
$n=100$    & updated\_experiment\_n\_100\_results.csv & img\_update\_100 \\
$n=200$    & updated\_experiment\_n\_200\_results.csv & img\_update\_200 \\
$n=300$    & updated\_experiment\_n\_300\_results\_1\_30.csv & img\_update\_updated\_300 \\
\bottomrule
\end{tabular}
\end{table}

\begin{table}[h!]
\footnotesize
\centering
\caption{Summary of Gurobi results for sample experiments (n=100)}
\label{tab:results}
\begin{tabular}{ccccc}
\toprule
experiment\_id & best\_cost & dt [s] & assignment & edges \\
\midrule
0 & 19.5360 & [...] & [...] & [...] \\
1 & 30.7337 & [...] & [...] & [...] \\
2 & 26.1106 & [...] & [...] & [...] \\
\bottomrule
\end{tabular}
\end{table}

\section{Experimental Results}

\begin{figure}[H]
    \centering
    \begin{subfigure}[b]{0.38\textwidth}
        \centering
        \includegraphics[width=\linewidth, trim=0 0 0 2.4cm, clip]{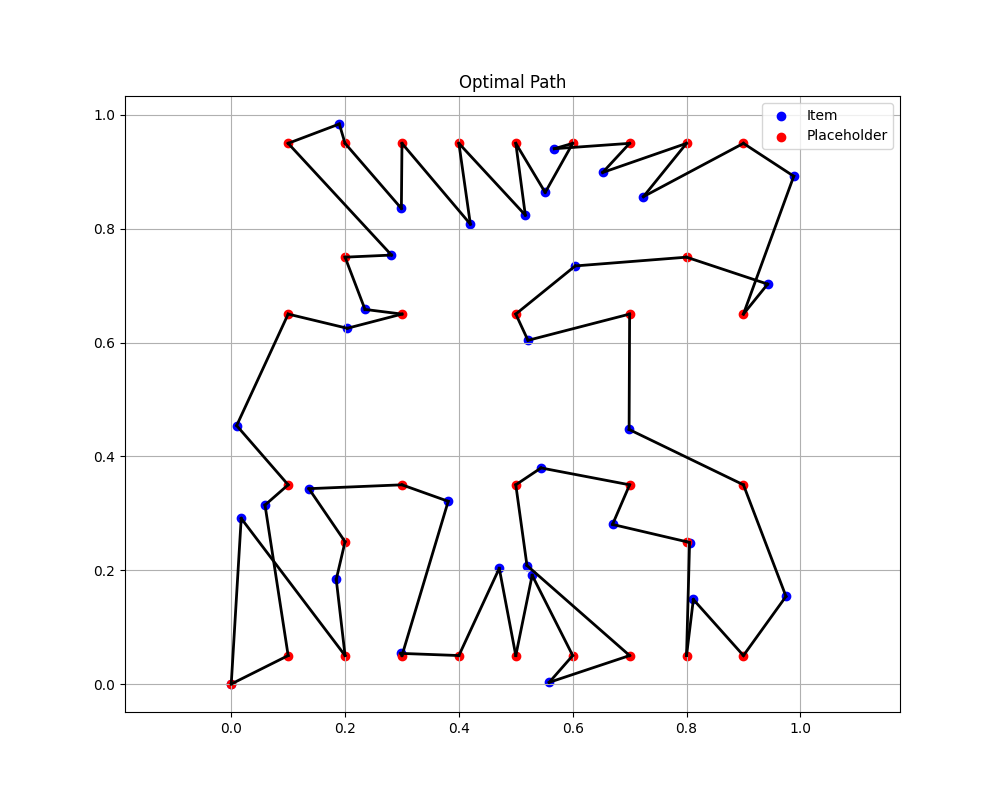}
        \caption{Experiment 1: Best path}
        \label{fig:xq32_1}
    \end{subfigure}
    \hspace{0.02\textwidth}
    \begin{subfigure}[b]{0.38\textwidth}
        \centering
        \includegraphics[width=\linewidth, trim=0 0 0 2.4cm, clip]{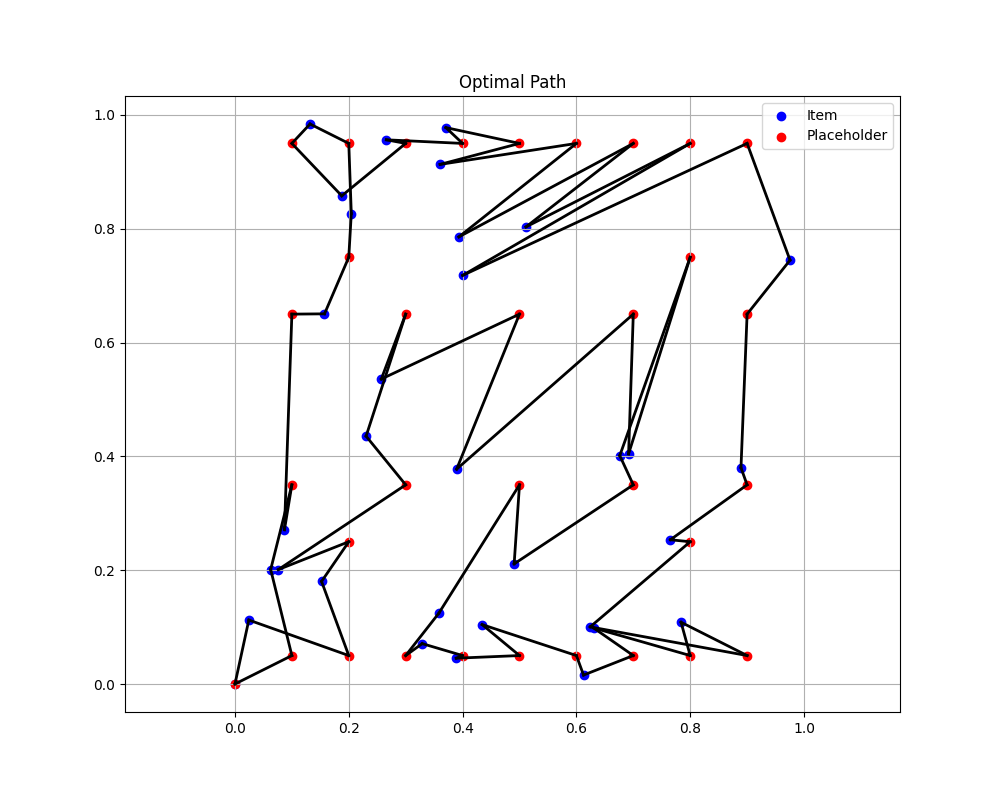}
        \caption{Experiment 2: Best path}
        \label{fig:xq32_2}
    \end{subfigure}

    \vspace{0.3em}

    \begin{subfigure}[b]{0.38\textwidth}
        \centering
        \includegraphics[width=\linewidth, trim=0 0 0 2.4cm, clip]{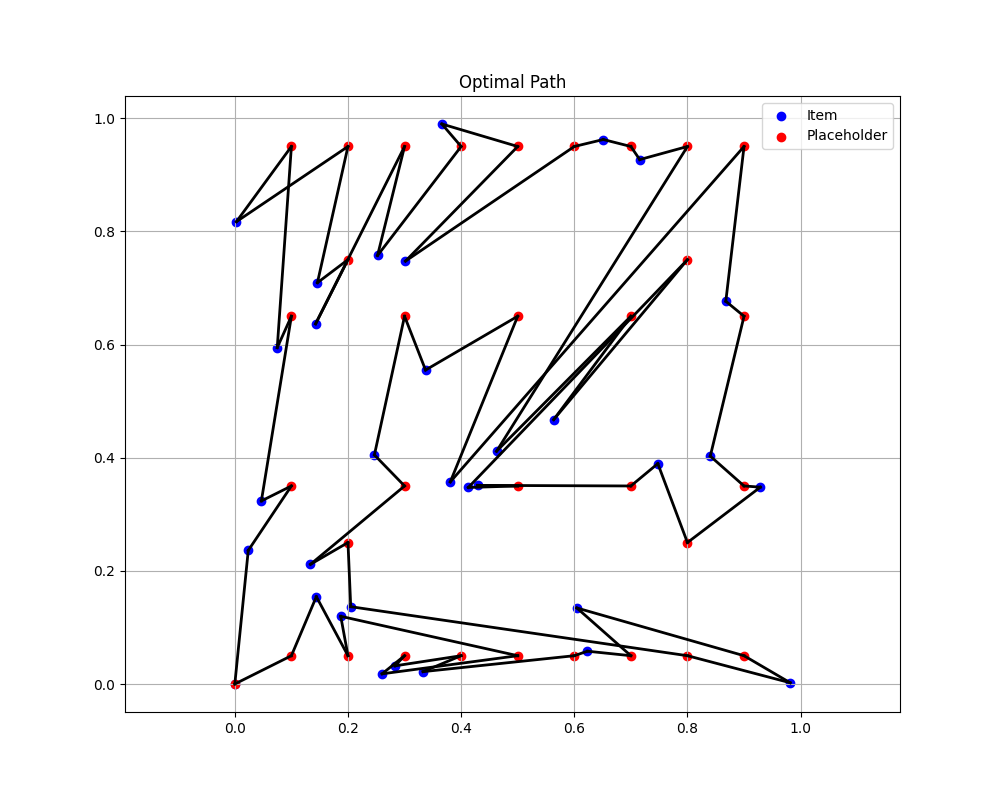}
        \caption{Experiment 3: Best path}
        \label{fig:xq32_3}
    \end{subfigure}
    \hspace{0.02\textwidth}
    \begin{subfigure}[b]{0.38\textwidth}
        \centering
        \includegraphics[width=\linewidth, trim=0 0 0 2.4cm, clip]{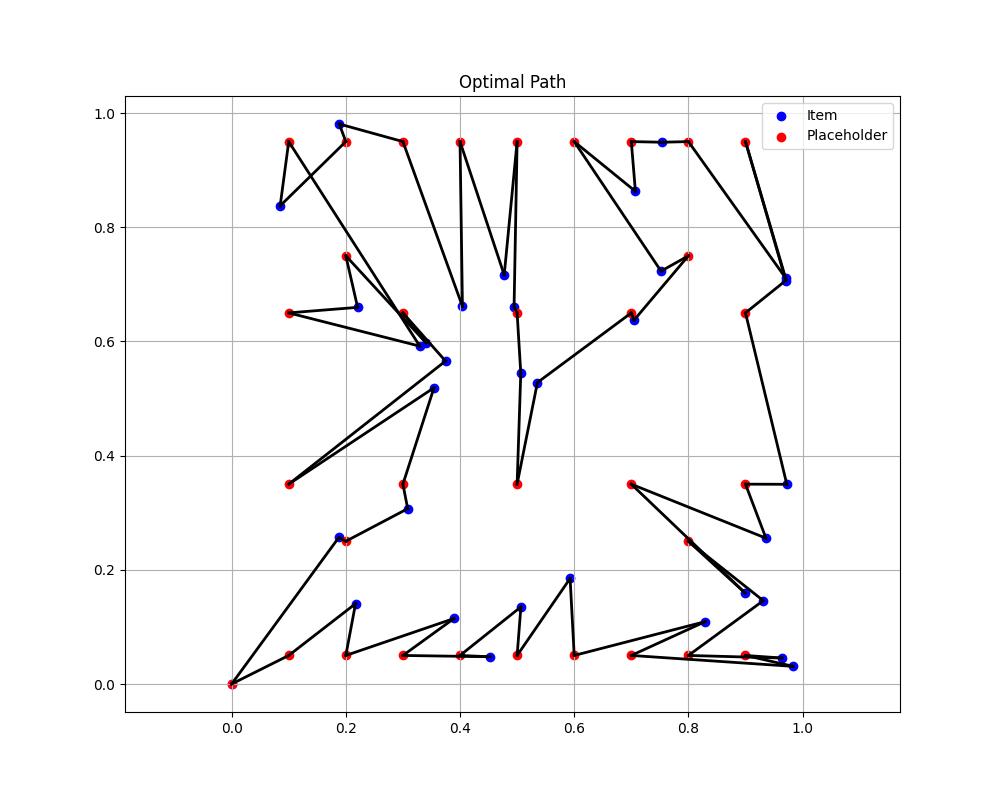}
        \caption{Experiment 4: Best path}
        \label{fig:xq32_4}
    \end{subfigure}

    \caption{First four sample results showing the computed optimal cycle for \(n=32\) (xq).}
    \label{fig:sample_results32xq}
\end{figure}

\begin{figure}[H]
    \centering
    \begin{subfigure}[b]{0.38\textwidth}
        \centering
        \includegraphics[width=\linewidth, trim=0 0 0 2.4cm, clip]{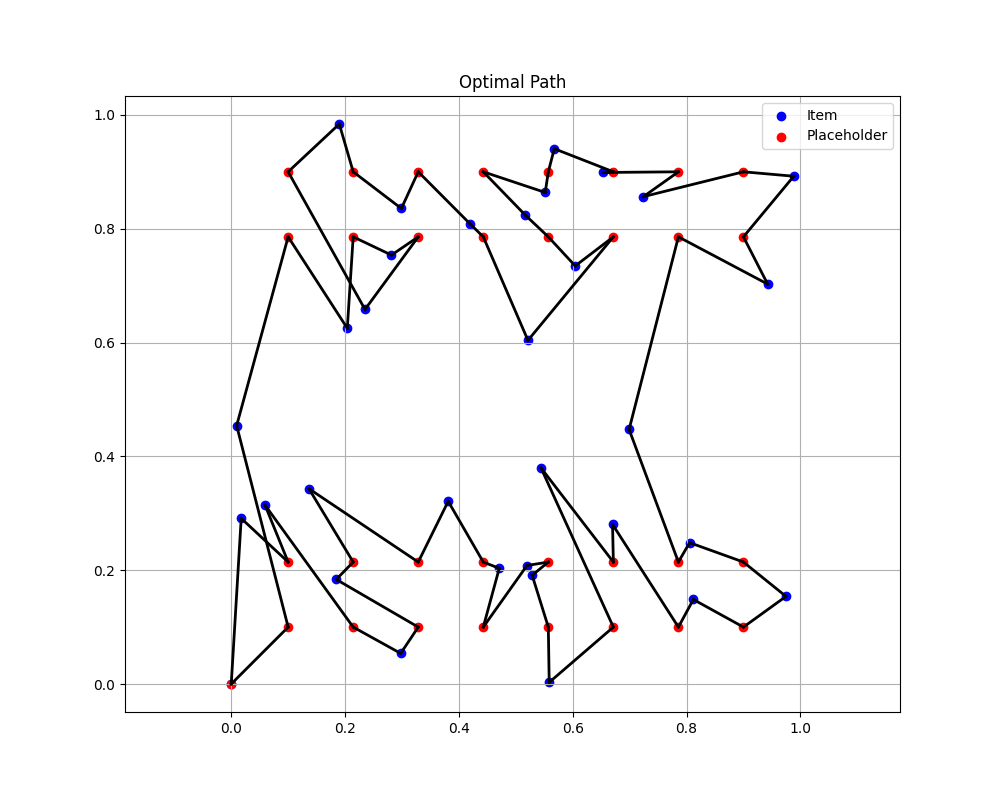}
        \caption{Experiment 1: Best path}
        \label{fig:ic32_1}
    \end{subfigure}
    \hspace{0.02\textwidth}
    \begin{subfigure}[b]{0.38\textwidth}
        \centering
        \includegraphics[width=\linewidth, trim=0 0 0 2.4cm, clip]{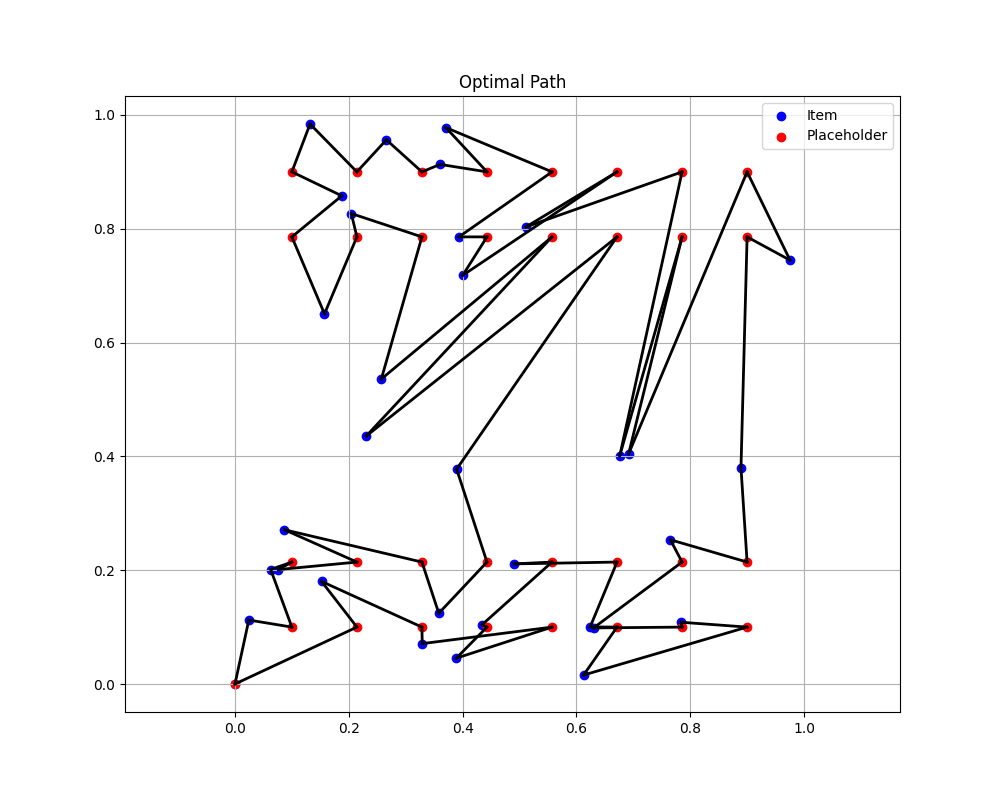}
        \caption{Experiment 2: Best path}
        \label{fig:ic32_2}
    \end{subfigure}

    \vspace{0.3em}

    \begin{subfigure}[b]{0.38\textwidth}
        \centering
        \includegraphics[width=\linewidth, trim=0 0 0 2.4cm, clip]{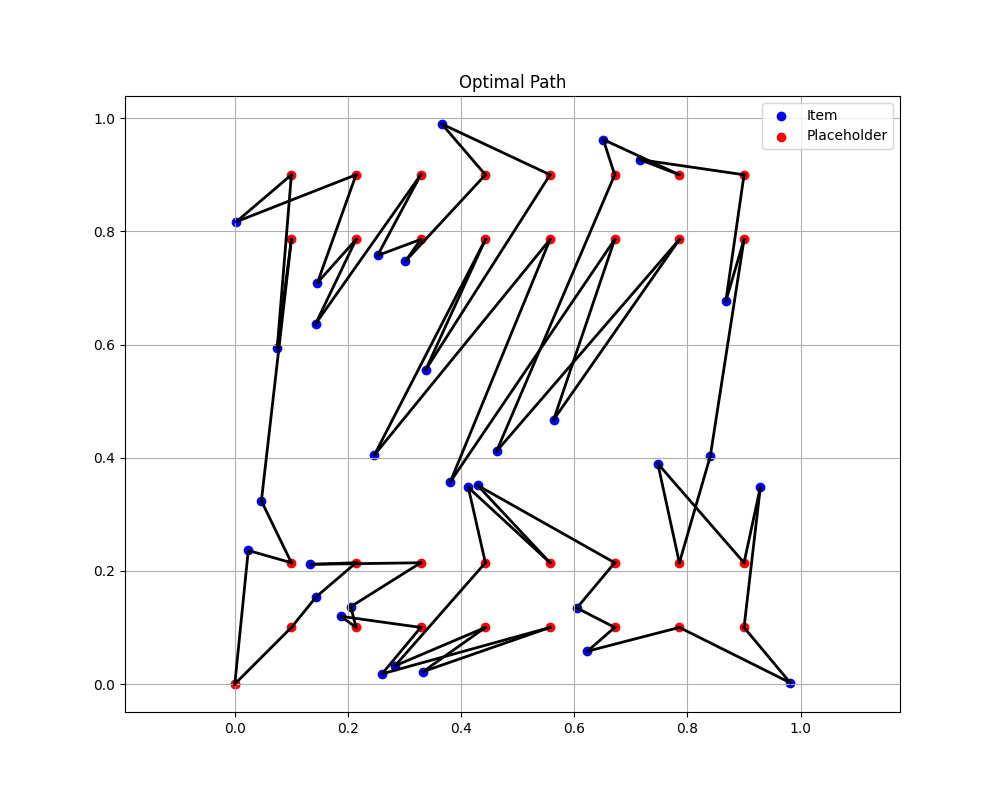}
        \caption{Experiment 3: Best path}
        \label{fig:ic32_3}
    \end{subfigure}
    \hspace{0.02\textwidth}
    \begin{subfigure}[b]{0.38\textwidth}
        \centering
        \includegraphics[width=\linewidth, trim=0 0 0 2.4cm, clip]{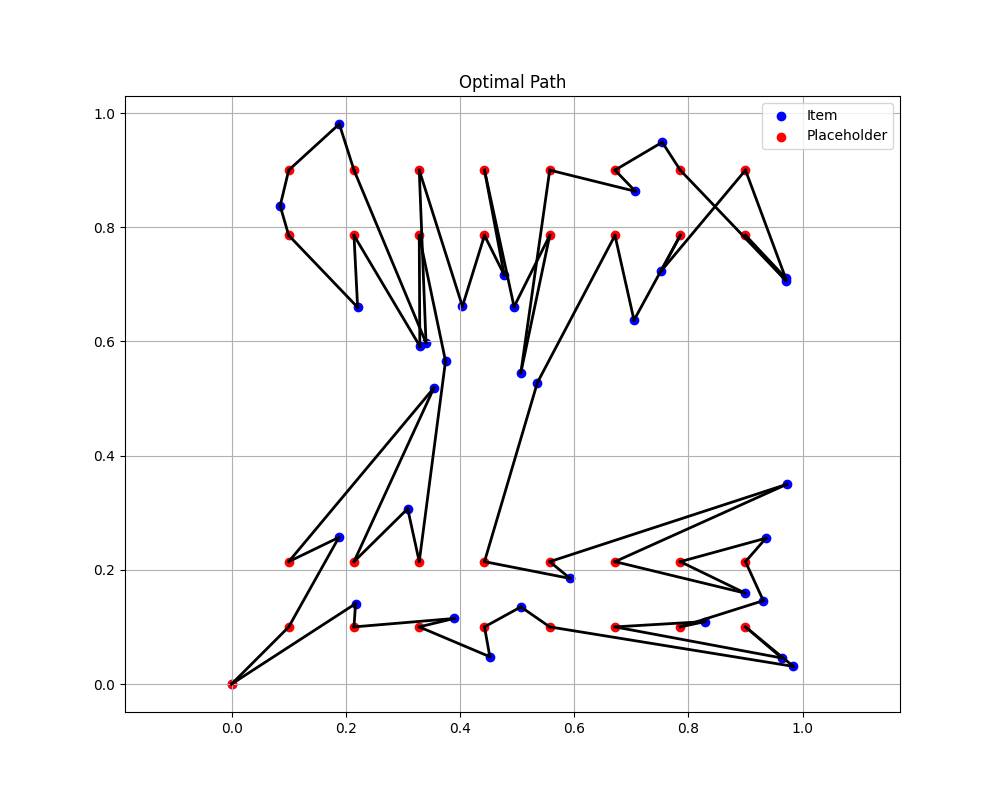}
        \caption{Experiment 4: Best path}
        \label{fig:ic32_4}
    \end{subfigure}

    \caption{First four sample results showing the computed optimal cycle for \(n=32\) (ic).}
    \label{fig:sample_results32ic}
\end{figure}

\begin{figure}[H]
    \centering
    \begin{subfigure}[b]{0.38\textwidth}
        \centering
        \includegraphics[width=\linewidth, trim=0 0 0 2.4cm, clip]{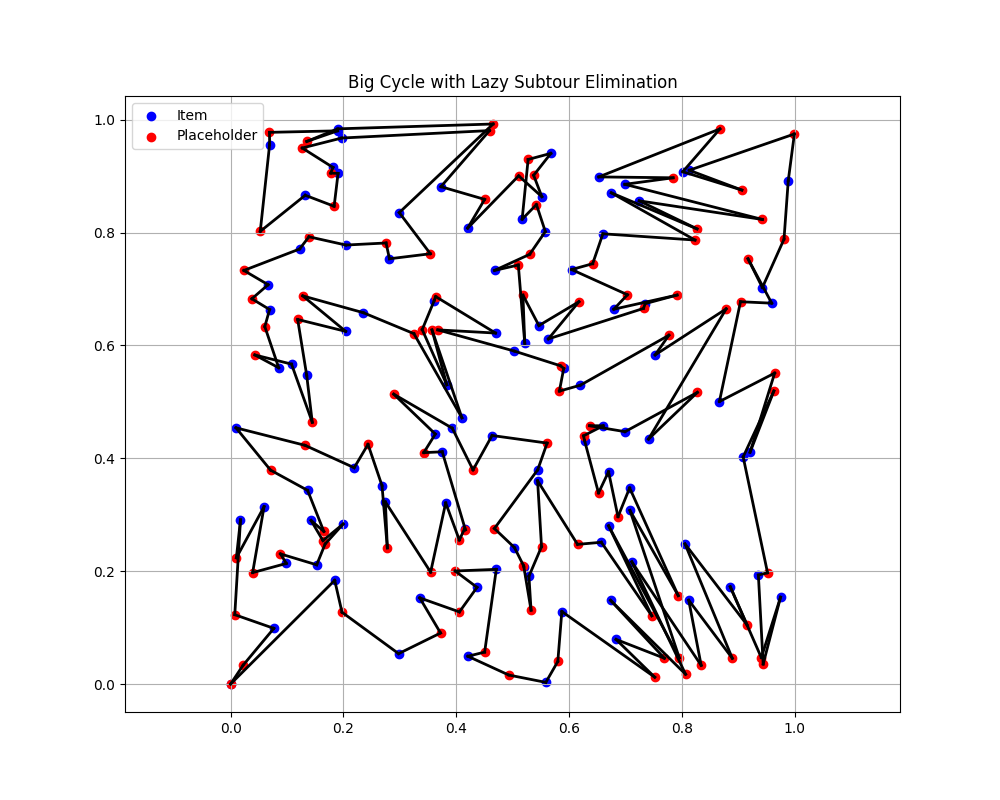}
        \caption{Experiment 1: Best path}
        \label{fig:100_1}
    \end{subfigure}
    \hspace{0.02\textwidth}
    \begin{subfigure}[b]{0.38\textwidth}
        \centering
        \includegraphics[width=\linewidth, trim=0 0 0 2.4cm, clip]{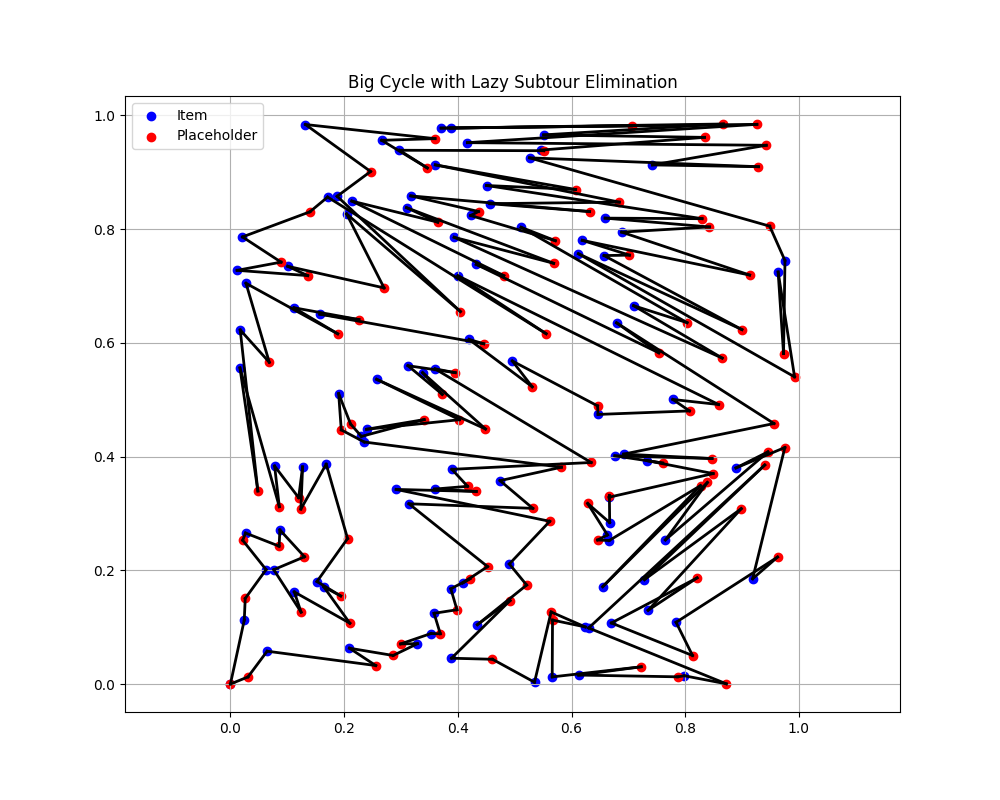}
        \caption{Experiment 2: Best path}
        \label{fig:100_2}
    \end{subfigure}

    \vspace{0.3em}

    \begin{subfigure}[b]{0.38\textwidth}
        \centering
        \includegraphics[width=\linewidth, trim=0 0 0 2.4cm, clip]{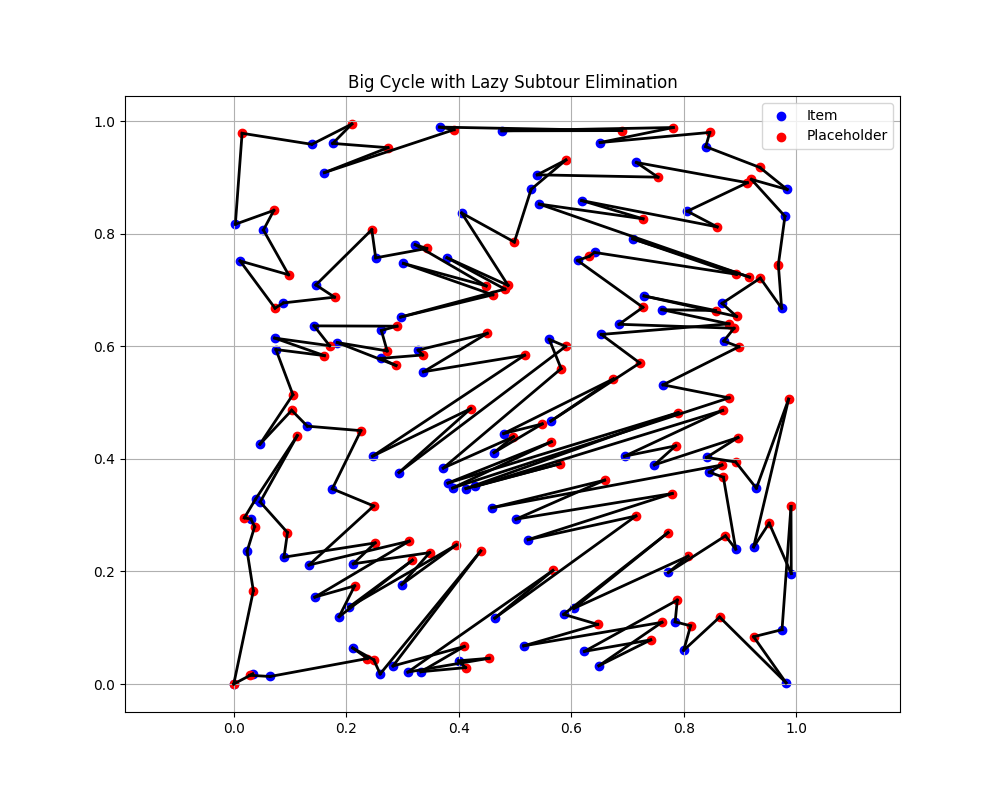}
        \caption{Experiment 3: Best path}
        \label{fig:100_3}
    \end{subfigure}
    \hspace{0.02\textwidth}
    \begin{subfigure}[b]{0.38\textwidth}
        \centering
        \includegraphics[width=\linewidth, trim=0 0 0 2.4cm, clip]{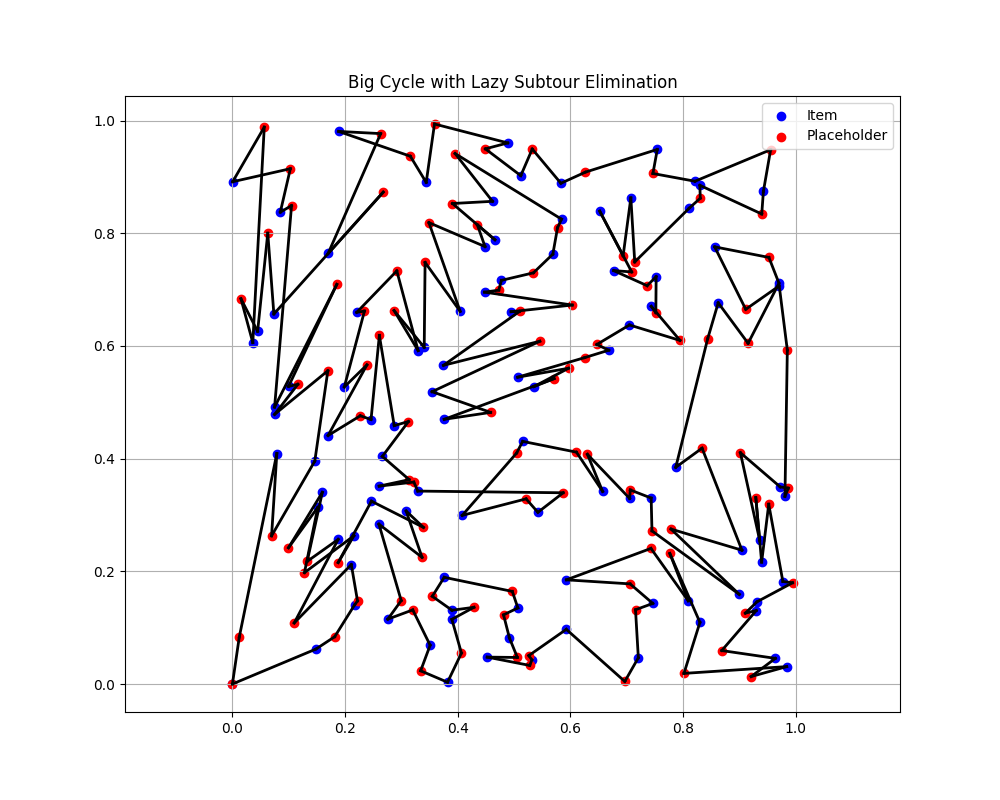}
        \caption{Experiment 4: Best path}
        \label{fig:100_4}
    \end{subfigure}

    \caption{First four sample results showing the computed optimal cycle for \(n=100\).}
    \label{fig:sample_results100}
\end{figure}

\begin{figure}[H]
    \centering
    \begin{subfigure}[b]{0.38\textwidth}
        \centering
        \includegraphics[width=\linewidth, trim=0 0 0 2.4cm, clip]{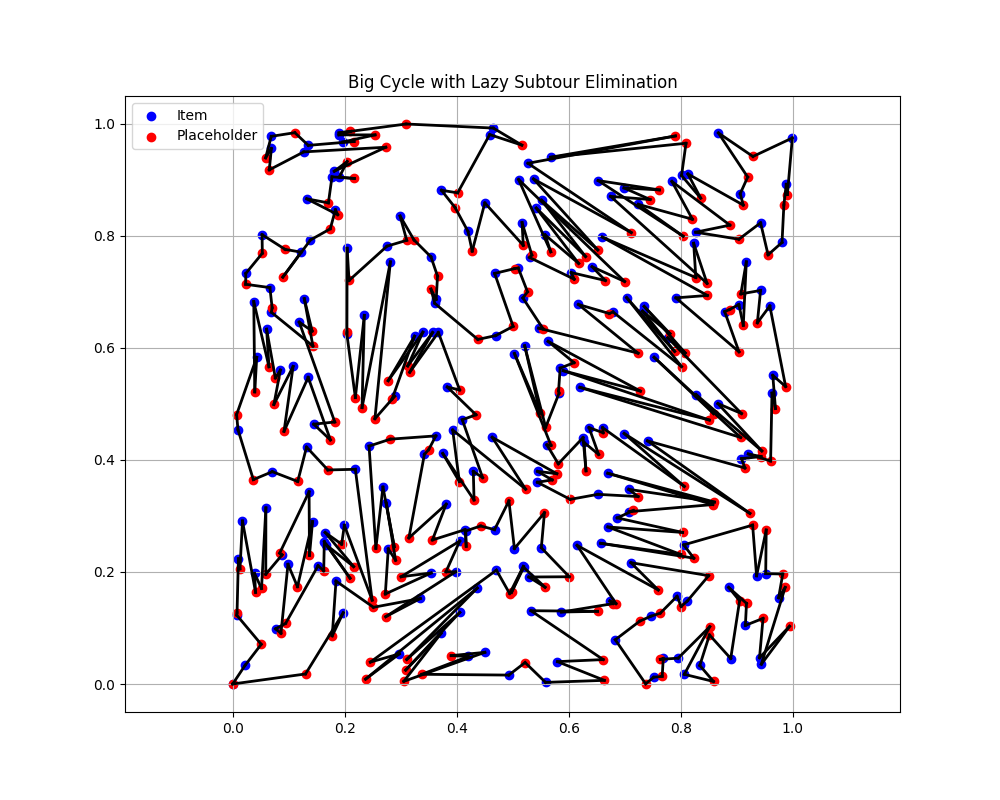}
        \caption{Experiment 1: Best path}
        \label{fig:200_1}
    \end{subfigure}
    \hspace{0.02\textwidth}
    \begin{subfigure}[b]{0.38\textwidth}
        \centering
        \includegraphics[width=\linewidth, trim=0 0 0 2.4cm, clip]{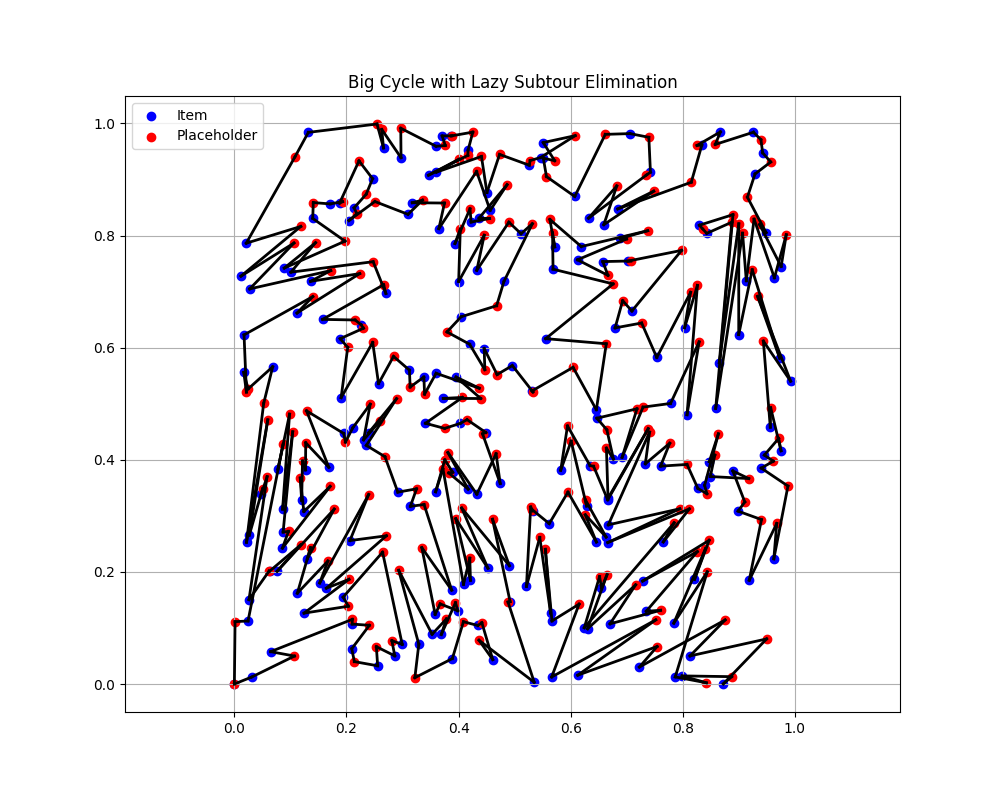}
        \caption{Experiment 2: Best path}
        \label{fig:200_2}
    \end{subfigure}

    \vspace{0.3em}

    \begin{subfigure}[b]{0.38\textwidth}
        \centering
        \includegraphics[width=\linewidth, trim=0 0 0 2.4cm, clip]{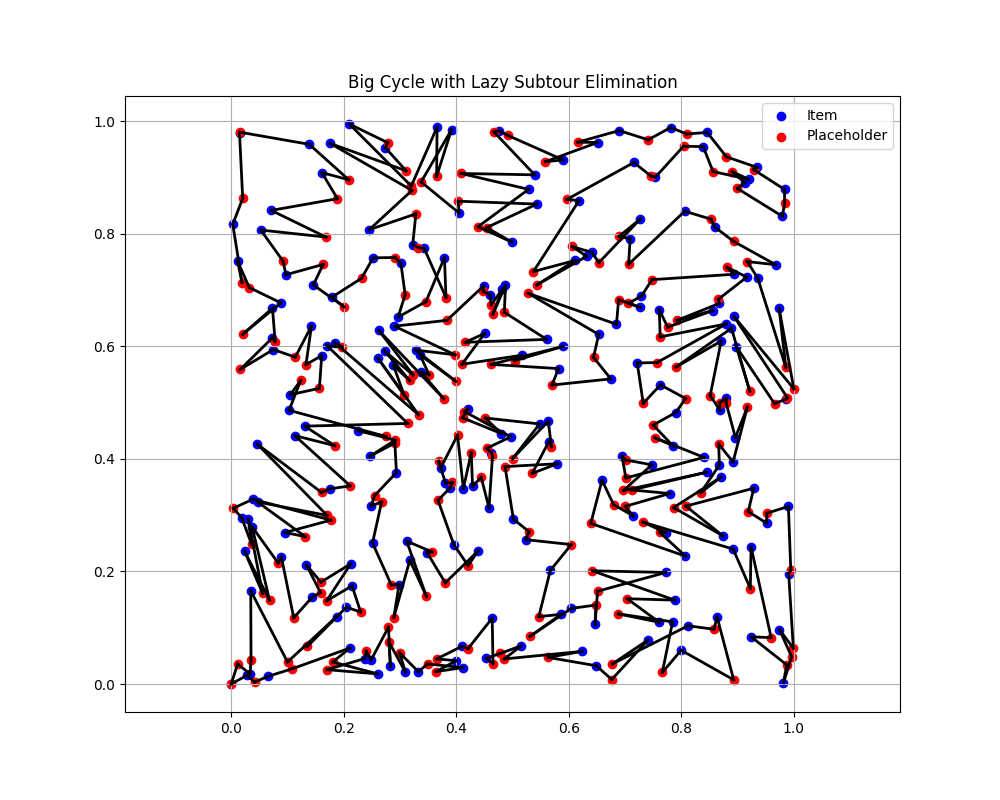}
        \caption{Experiment 3: Best path}
        \label{fig:200_3}
    \end{subfigure}
    \hspace{0.02\textwidth}
    \begin{subfigure}[b]{0.38\textwidth}
        \centering
        \includegraphics[width=\linewidth, trim=0 0 0 2.4cm, clip]{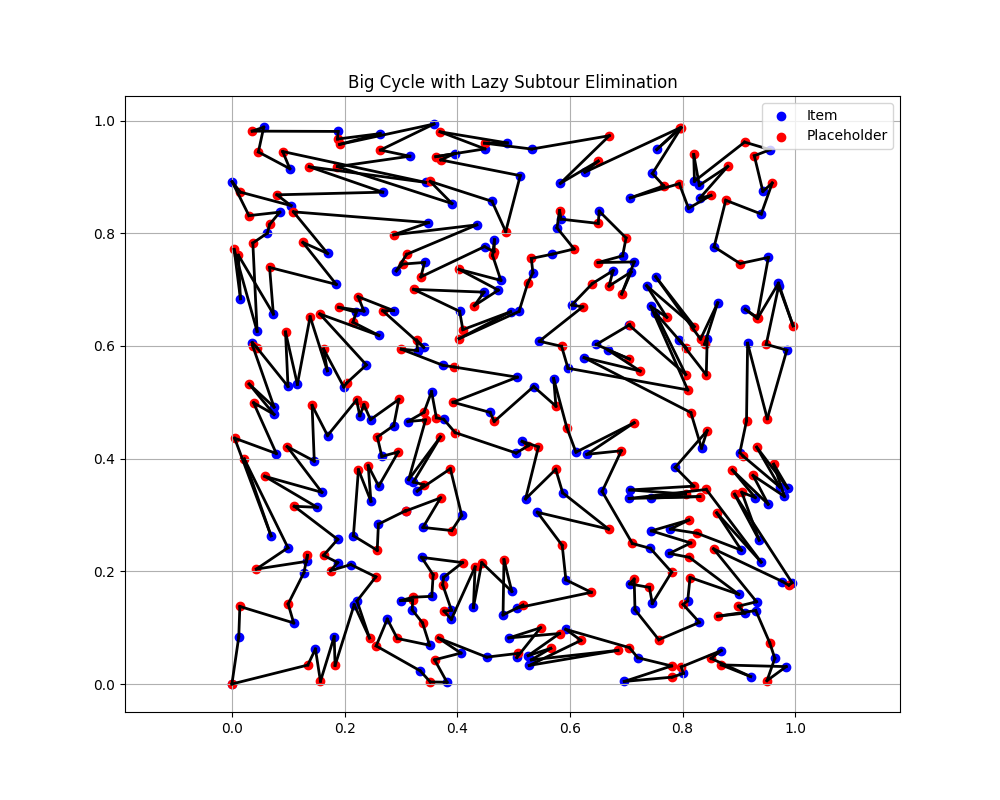}
        \caption{Experiment 4: Best path}
        \label{fig:200_4}
    \end{subfigure}

    \caption{First four sample results showing the computed optimal cycle for \(n=200\).}
    \label{fig:sample_results200}
\end{figure}

Samples of the results are shown in Figures~\ref{fig:sample_results32xq},~\ref{fig:sample_results32ic},~\ref{fig:sample_results100},~\ref{fig:sample_results200}, and summarized in Table~\ref{tab:merged_results_updated}. The complete dataset of experimental results is available in the accompanying GitHub repository for full reproducibility. As evidenced in Table~\ref{tab:merged_results_updated}, the exact solver performs efficiently for small instances, consistently returning optimal solutions in under a second for problem sizes up to $n = 100$. However, as the problem size increases — particularly for $n = 300$, the computational time reach one minutes. When n reach 1000, the solver often fails to provide a valid solution.

These empirical findings highlight the inherent \textbf{scalability limitations of exact Mixed-Integer Programming (MIP) formulations} for combinatorial problems of this nature.

Such performance trends are consistent with the theoretical complexity of the problem: the formulation generalizes the \textit{Traveling Salesman Problem (TSP)}, which is well-known to be NP-hard. The worst-case computational complexity of branch-and-cut solvers applied to MIP formulations typically grows \emph{exponentially with the number of binary variables}, which in this case is $m = \Theta(n^2)$, due to pairwise interactions between nodes.

These results underscore a fundamental computational barrier: while exact methods are valuable for small instances or benchmarking purposes, their exponential growth in runtime renders them impractical for large-scale instances. Hence, developing and analyzing \emph{approximation algorithms or heuristics} that deliver \textbf{near-optimal solutions} with significantly lower computational cost becomes not only practical but necessary. Such approaches can maintain high-quality solutions while scaling efficiently to larger values of $n$.

\begin{table}[h!]
\footnotesize
\centering
\caption{Merged experimental results for $n=32$ XQ, $n=32$ Chess, $n=100$, $n=200$, and $n=300$ placeholders.}
\label{tab:merged_results_updated}
\resizebox{\textwidth}{!}{%
\begin{tabular}{c c c|c c|c c|c c|c c}
\toprule
\multicolumn{3}{c|}{$n=32$ XQ} & 
\multicolumn{2}{c|}{$n=32$ Chess} & 
\multicolumn{2}{c|}{$n=100$} & 
\multicolumn{2}{c|}{$n=200$} & 
\multicolumn{2}{c}{$n=300$} \\
\cmidrule(lr){1-3} \cmidrule(lr){4-5} \cmidrule(lr){6-7} \cmidrule(lr){8-9} \cmidrule(lr){10-11}
Item & Best Cost & Time (s) &
Best Cost & Time (s) &
Best Cost & Time (s) &
Best Cost & Time (s) &
Best Cost & Time (s) \\
\midrule
1  & 8.9811 & 0.1749 & 8.6045 & 0.0640 & 19.5360 & 0.5980 & 32.5257 & 1.8281 & 42.0017 & 76.0503 \\
2  & 11.1315 & 0.0184 & 11.4452 & 0.0204 & 30.7337 & 0.7149 & 29.7790 & 8.3149 & 33.1289 & 25.6637 \\
3  & 12.5431 & 0.2018 & 13.5874 & 0.0154 & 26.1106 & 0.8715 & 26.9493 & 4.2116 & 34.3303 & 44.0879 \\
4  & 10.6235 & 0.0195 & 11.2942 & 0.0401 & 19.0954 & 0.9859 & 27.4590 & 6.2985 & 35.1968 & 28.5358 \\
5  & 10.9402 & 0.0450 & 10.8882 & 0.0494 & 18.5475 & 2.0169 & 28.5458 & 14.5061 & 33.8118 & 77.8469 \\
6  & 8.5998  & 0.0611 & 9.5426  & 0.0869 & 19.2728 & 1.5554 & 32.5525 & 6.9940 & 37.4785 & 84.7612 \\
7  & 10.8217 & 0.0149 & 11.3253 & 0.0717 & 20.7782 & 0.8501 & 38.1054 & 8.7085 & 33.9530 & 26.5168 \\
8  & 9.3054  & 0.1165 & 9.8737  & 0.0133 & 25.1340 & 0.2294 & 38.0598 & 4.1589 & 40.6682 & 28.3229 \\
9  & 10.6951 & 0.0742 & 11.3153 & 0.0821 & 19.9261 & 0.2498 & 31.5435 & 10.0047 & 36.6766 & 51.3296 \\
10 & 11.4083 & 0.0331 & 12.3702 & 0.0123 & 20.6597 & 2.6301 & 28.0285 & 11.0655 & 38.6581 & 83.7323 \\
\midrule
\textbf{Average} & -- & \textbf{0.0758} & -- & \textbf{0.0452} & -- & \textbf{0.9709} & -- & \textbf{7.6993} & -- & \textbf{52.2849} \\
\bottomrule
\end{tabular}%
}
\end{table}

\section{Conclusion}
\label{sec:conclusion}

In this work, we addressed a joint assignment–routing optimization problem, where item-to-placeholder pairings and a Hamiltonian tour are determined simultaneously to minimize total travel cost. We formulated the problem as a compact mixed-integer program (MIP) and solved it using Gurobi, employing a cutting-plane approach for subtour elimination to ensure tour feasibility. Many pickup-placing with routing problems, for example, arranging chess game (with different type of pieces) and packing food/ toys with different item patterns, can be solved with such method with addition assignment constraints without increasing the complexity of the problem.

Our experimental analysis confirms that the exact MIP solver performs well for moderate-sized instances (e.g., $n = 30$), typically producing optimal solutions in fractions of a second. However, as the problem size increases, computational requirements grow rapidly: for $n = 300$, solving times reach 1 minute; for $n = 1000$, solver often fails to work. These results highlight the scalability limitations of exact MIP-based methods in practice.

This study underscores the need for more scalable approaches — such as approximation algorithms, heuristics, or hybrid learning–search methods — to tackle larger problem sizes efficiently while maintaining high solution quality. Our publicly released dataset and benchmark results support reproducibility and offer a foundation for evaluating future algorithmic developments. By providing this benchmark, we aim to enable systematic comparison of solution quality, runtime, and empirical scaling across methods in future research.

\paragraph{Use of Generative AI.}
This manuscript involved the use of generative AI tools (ChatGPT-4, OpenAI) during the drafting process. These tools were employed to assist with language refinement, equation formatting, and improving the clarity of technical explanations. All scientific content, methods, experimental results, and conclusions were developed and validated by the authors. Final manuscript versions were thoroughly reviewed to ensure accuracy and originality.

\bibliographystyle{unsrt}  
\bibliography{references}  

\begin{thebibliography}{10}

\bibitem{hernandez2004}
H.~Hern{\'a}ndez-P{\'e}rez and J.~J. Salazar-Gonz{\'a}lez.
\newblock Heuristics for the one-commodity pickup-and-delivery traveling
  salesman problem.
\newblock {\em Transportation Science}, 38(2):245--255, 2004.

\bibitem{hernandez2007}
H.~Hern{\'a}ndez-P{\'e}rez and J.~J. Salazar-Gonz{\'a}lez.
\newblock The one-commodity pickup-and-delivery traveling salesman problem:
  Inequalities and algorithms.
\newblock {\em Networks}, 50(4):258--272, 2007.

\bibitem{treleaven2012}
K.~Treleaven, M.~Pavone, and E.~Frazzoli.
\newblock Asymptotically optimal algorithms for pickup and delivery problems
  with application to large-scale transportation systems.
\newblock {\em arXiv preprint}, 2012.

\bibitem{kuhn1955hungarian}
Harold~W Kuhn.
\newblock The hungarian method for the assignment problem.
\newblock {\em Naval Research Logistics Quarterly}, 2(1-2):83--97, 1955.

\bibitem{dantzig1954solution}
George~B Dantzig, Ray Fulkerson, and Selmer~M Johnson.
\newblock Solution of a large-scale traveling-salesman problem.
\newblock {\em Operations Research}, 2(4):393--410, 1954.

\bibitem{lawler1985traveling}
Eugene~L Lawler, Jan~Karel Lenstra, Alexander H~G Rinnooy~Kan, and David~B
  Shmoys.
\newblock {\em The Traveling Salesman Problem: A Guided Tour of Combinatorial
  Optimization}.
\newblock Wiley, 1985.

\bibitem{applegate2006traveling}
David~L Applegate, Robert~E Bixby, Vasek Chv{\'a}tal, and William~J Cook.
\newblock {\em The Traveling Salesman Problem: A Computational Study}.
\newblock Princeton University Press, 2006.

\bibitem{gurobi2024manual}
{Gurobi Optimization, LLC}.
\newblock {\em Gurobi Optimizer Reference Manual}, 2024.

\bibitem{kelley1960cutting}
James~E Kelley, Jr.
\newblock The cutting-plane method for solving convex programs.
\newblock {\em Journal of the society for Industrial and Applied Mathematics},
  8(4):703--712, 1960.

\bibitem{westerlund1995extended}
Tapio Westerlund and Frank Pettersson.
\newblock An extended cutting plane method for solving convex minlp problems.
\newblock {\em Computers \& chemical engineering}, 19:131--136, 1995.

\bibitem{basu2020complexitybranchandboundcuttingplanes}
Amitabh Basu, Michele Conforti, Marco~Di Summa, and Hongyi Jiang.
\newblock Complexity of branch-and-bound and cutting planes in mixed-integer
  optimization, 2020.

\bibitem{karmarkar1984new}
Narendra Karmarkar.
\newblock A new polynomial-time algorithm for linear programming.
\newblock In {\em Proceedings of the sixteenth annual ACM symposium on Theory
  of computing}, pages 302--311, 1984.

\bibitem{gurobi2025mipprimer}
Gurobi Optimization.
\newblock Mixed-integer programming (mip / milp): An introduction to the
  basics, 2025.
\newblock Accessed: 2025-10-13.

\end{thebibliography}

\end{document}